\documentclass{ws-ijmpd}

\begin{document}

\markboth{Melia}
{The Cosmological Spacetime}

\catchline{}{}{}{}{}
%
%%%%%%%%%%%%%%%%%%%%%%%%%%%%%%%%%%%%%%%%%%%%%%%%%%%%%%%%%%%%%%%%%%%%

\title{The Cosmological Spacetime}

\author{FULVIO MELIA
\and
MAJD ABDELQADER}

\address{Department of Physics and Steward Observatory,\\
The University of Arizona,\\
Tucson, AZ 85721, USA\\
melia@physics.arizona.edu}

\maketitle

\begin{history}
\received{4 June 2009}
\revised{17 July 2009}
\accepted{}
%\comby{Managing Editor}
\end{history}

\begin{abstract}
We present here the transformations required to recast the Robertson-Walker metric
and Friedmann-Robertson-Walker equations in terms of observer-depen\-dent coordinates
for several commonly assumed cosmologies. The overriding motivation is the derivation
of explicit expressions for the radius $R_{\rm h}$ of our cosmic horizon in terms of
measurable quantities for each of the cases we consider. We show that the cosmological
time $dt$ diverges for any finite interval $ds$ associated with a process at $R\rightarrow R_{\rm h}$,
which therefore represents a physical limit to our observations. This is a key component
required for a complete interpretation of the data, particularly as they pertain to the
nature of dark energy. With these results, we affirm the conclusion drawn in our earlier
work that the identification of dark energy as a cosmological constant does not appear to
be consistent with the data.
\end{abstract}

\keywords{cosmology; dark energy; gravitation.}

\section{Introduction}	
Standard cosmology is based on the Robertson-Walker (RW) metric for a spatially homogeneous
and isotropic three-dimensional space, expanding or contracting as a function of time:
\begin{equation}
ds^2=c^2\,dt^2-a^2(t)[dr^2(1-kr^2)^{-1}+
r^2(d\theta^2+\sin^2\theta\,d\phi^2)]\;.
\end{equation}
\vskip 0.1in\noindent
In the coordinates used for this metric, $t$ is the cosmic time, measured by a
comoving observer (and is the same everywhere), $a(t)$ is the expansion factor, and $r$ is an
appropriately scaled radial coordinate in the comoving frame. The geometric factor $k$ is $+1$
for a closed universe, $0$ for a flat universe, and $-1$ for an open universe.

The expansion of the universe is calculated from the Friedmann-Robertson-Walker (FRW)
differential equations of motion,
\begin{equation}
H^2\equiv\left(\frac{\dot a}{a}\right)^2=\frac{8\pi G}{3c^2}\rho-\frac{kc^2}{a^2}\;,
\end{equation}
\begin{equation}
\frac{\ddot a}{a}=-\frac{4\pi G}{3c^2}(\rho+3p)\;,
\end{equation}
\begin{equation}
\dot\rho=-3H(\rho+p)\;,
\end{equation}
derived from an application of the RW metric to the Einstein field equations. Here,
an overdot denotes a derivative with respect to $t$, and $\rho$ and $p$
represent, respectively, the total energy density and total pressure. Often, the
latter is written as $p=w\rho$, and $w$ is then used to characterize
the expansion properties of the medium. For example, $w\approx 0$ for (visible
and dark) matter, $+1/3$ for radiation, and $-1$ for a pure cosmological constant
(though the actual value of $w$ is not restricted to just these values).

In previous papers,\cite{M07,M08} we demonstrated the usefulness of expressing
the RW metric in terms of an observer-dependent coordinate $R=a(t)r$, which
explicitly reveals the dependence of the observed intervals of distance, $dR$,
and time on the curvature induced by the mass-energy content between the observer
and $R$; in the metric, this effect is represented by the proximity of the
{\it physical} radius $R$ to the cosmic horizon $R_{\rm h}$, defined by the relation
\begin{equation}
{2GM(R_{\rm h})\over c^2}=R_{\rm h}\;.
\end{equation}
In this expression, $M(R_{\rm h})$ is the mass enclosed within $R_{\rm h}$. In terms
of $\rho$, we may also write $R_{\rm h}=(3c^4/8\pi G\rho)^{1/2}$ or, more simply,
$R_{\rm h}=c/H(t)$ in a flat universe. This is the radius at which a sphere encloses
sufficient mass-energy to create a significant time dilation for an observer at the
surface relative to the origin of the coordinates.

When the RW metric is written in terms of $R$, the presence of $R_{\rm h}$ alters
the intervals of time we measure progressively
more and more as $R\rightarrow R_{\rm h}$. And since the gravitational time dilation
becomes infinite near $R_{\rm h}$, it is physically impossible for us to see any process
occurring beyond this distance. What this means, of course, is that light emitted
beyond $R_{\rm h}$ is infinitely redshifted and therefore carries no signal.

The motivation for the introduction of these new coordinates was the set of
recent observations pointing to a cosmic-horizon radius $R_{\rm h}(t_0)\approx
ct_0$, where $t_0$ is the inferred current age of the universe.\cite{M07} (Throughout
this paper, a subscript $0$ denotes quantities measured at the present time.) 
This inference is based on precision measurements\cite{S03} of the CMB radiation,
which indicate that the universe is extremely flat (i.e., $k=0$). Thus $\rho$ is at 
(or very near) the ``critical" value $\rho_0\equiv 3c^2H_0^2/8\pi G$. The Hubble 
Space Telescope Key Project\cite{M00} on the extragalactic distance scale has measured 
$H$ with unprecedented accuracy, finding a current value $H_0\equiv H(t_0)=71\pm6$ 
km s$^{-1}$ Mpc$^{-1}$. It is straightforward to show that with this $H_0$, 
$R_0\approx ct_0$ (specifically, $13.4$ versus $13.7$ billion lightyears).
But this {\it empirical} result is very peculiar because the FRW equations predict 
that $\dot R_{\rm h}=c$ only for the very special equation of state $w=-1/3$ which, 
however, does not appear to be consistent with any of the known constituents of the 
universe, including a cosmological constant with $w=w_\Lambda\equiv -1$. 

As a brief aside, it is interesting to note that de Sitter's own metric (for a
universe containing only a cosmological constant) was first written in terms of
observer-dependent coordinates,\cite{d17} though
this is no longer widely known, and everyone now uses the form of
the RW metric written in terms of $r$ and $t$ only. Of course, de Sitter's work
was completed prior to the introduction of the comoving coordinates $(ct,r,\theta,\phi)$
several years later. A physical motivation for how one might arrive at de Sitter's
metric from an application of Schwarzschild's solution to a uniform
infinite medium may be understood in terms of Birkhoff's theorem and its 
corollary.\cite{M07,B23}

Given the significance of the fact that $R_{\rm h}(t_0)\approx ct_0$, our
earlier discussion on this topic\cite{M07,M08} was largely based on a simple
cosmological spacetime with $w=-1/3$. The condition $\dot R_{\rm h}=c$ is then
always true, guaranteeing the equality (or near equality) of $R_{\rm h}$ and $ct$.
However, if the equation of state is time dependent, $w$ need not be exactly
$-1/3$ in order to achieve the result $R_{\rm h}(t_0)=ct_0$ in the current universe.
Indeed, Type Ia supernova data seem to suggest a value of $w$ smaller than this,\cite{R98,P99} 
which motivates us to consider a broader
range of constituents in the Universe in order to fully interrogate the available data.

The purpose of this paper is to present the transformations required to recast the RW
metric and FRW equations for a flat universe ($k=0$) into observer-dependent coordinates
for several commonly assumed cosmologies. The overriding motivation is the derivation of
an explicit expression for $R_{\rm h}$ in terms of measurable quantities for each of the
cases we consider. As we shall see, this is a key component required for a complete
interpretation of the observations, particularly as they pertain to the nature of dark
energy. In the process, we shall affirm the conclusion drawn earlier\cite{M07,M08}
that dark energy is probably not a cosmological constant.

\section{General Coordinate Transformation}
In order to write the RW metric in terms of observer-dependent coordinates,
we begin with the physical radius
\begin{equation}
R\equiv a(t)\,r\;.
\end{equation}
However, it will become evident later that to simplify the mathematical transformations,
it is convenient to recast Equation (1) using a new function $f(t)$, where
\begin{equation}
a(t)=e^{f(t)}\;.
\end{equation}
In that case,
\begin{equation}
ds^2=c^2\, dt^2- e^{2f(t)}\left[ dr^2+r^2\, d\Omega^2\right]\;,
\end{equation}
where
\begin{equation}
d\Omega^2\equiv d\theta^2+\sin^2\theta\,d\phi^2\;.
\end{equation}
Now,
\begin{equation}
r=R\, e^{-f}\;,
\end{equation}
and therefore
\begin{equation}
dr= e^{-f} \left[ dR-R \dot{f}\, dt\right]\;,
\end{equation}
so that
\begin{equation}
dr^2 = e^{-2f} \left[ dR^2+ \left( \frac{R\dot{f}}{c}\right)^2\, c^2\,dt^2 -2 
\left(\frac{R\dot{f}}{c} \right)c\,dt\,dR \right]\;.
\end{equation}

Collecting terms, we can now write the metric as 
\begin{equation}
ds^2 = \left[1-\left(\frac{R\dot{f}}{c} \right)^2 \right]\,c^2\,dt^2 +
2 \left(\frac{R\dot{f}}{c}\right) c\,dt\,dR-dR^2 -R^2\,d\Omega^2\;,
\end{equation}
whereupon, completing the square, we see that
\begin{equation}
ds^2= \Phi\left[c\,dt + \left(\frac{R\dot{f}}{c} \right)\Phi^{-1} 
dR  \right]^2 - \Phi^{-1}{dR^2}-R^2\,d\Omega^2\;.
\end{equation}
For convenience we have also defined the quantity
\begin{equation}
\Phi\equiv 1-\left(\frac{R\dot{f}}{c} \right)^2\;,
\end{equation}
which appears frequently in the metric coefficients. We can easily see that
the radius of the cosmic horizon for the observer at the origin is
\begin{equation}
R_{\rm h}\equiv c/\dot{f}
\end{equation}
so that, in effect, the function
\begin{equation}
\Phi\equiv 1-\left(\frac{R}{R_{\rm h}} \right)^2
\end{equation}
signals the dependence of the metric on the proximity of the observation radius
$R$ to the maximum observable distance $R_{\rm h}$.
Equation~(14) thus becomes
\begin{equation}
ds^2= \Phi\left[c\,dt + \left(\frac{R}{R_{\rm h}} \right)\Phi^{-1} 
dR  \right]^2 - \Phi^{-1}{dR^2}-R^2\,d\Omega^2\;.
\end{equation}

This metric is already in a form we can use to examine the behavior
of the cosmological spacetime at any radius $R$. However, sometimes 
(e.g., as in the de Sitter metric) it may be useful to ``complete" 
the transformation by introducing a new time coordinate $T$, such that
\begin{equation}
\Phi^{-1/2}\frac{c\,dT}{\eta(t,R)}\equiv \Phi^{1/2}c\,dt +
\left(\frac{R}{R_{\rm h}} \right)\Phi^{-1/2} \, dR\;,
\label{ansatz}
\end{equation}
where $\eta(t,R)$ is an integrating factor selected to guarantee that $dT$ is an
exact differential.\cite{M07}  There are possibly an infinite number of solutions
for $\eta(t,R)$, but only one is dimensionless (ensuring that $T$ has dimensions of 
time), while completely diagonalizing the metric. With this substitution,
\begin{equation}
ds^2=\Phi^{-1}\frac{c^2\, dT^2}{\eta^2(t,R)}- \Phi^{-1}\,{dR^2}
-R^2\,d\Omega^2\;.
\label{finalmetric}
\end{equation}
But in order for $dT$ to be an exact differential, $T$ must satisfy the following condition:
\begin{equation}
\frac{\partial^2 \, T}{\partial R \, \partial t}=\frac{\partial^2\, T}{\partial t \, \partial R}\;,
\end{equation}
which means that $\eta(t,R)$ must be a solution to the equation
\begin{equation}
\frac{\partial}{\partial R} \left[\Phi\;\eta(t,R) \right]
= \frac{\partial}{\partial t} \left[\left(\frac{R}{cR_{\rm h}} \right)\,
\eta(t,R) \right]\;.
\label{mainDE}
\end{equation}

As we shall learn shortly, however, the coordinate $T$ has only limited
applicability because it is rarely possible to integrate $dT$ starting
from the big bang at $t=0$ to the present. 

Let us now examine the behavior of the interval $ds$ connecting any
arbitrary pair of spacetime events at $R$. For an interval produced at 
$R$ by the advancement of time only (with $dR=d\Omega=0$), Equation~(18) gives 
\begin{equation}
ds^2=\Phi\,c^2\,dt^2\;.
\end{equation}
Thus, if we were to make a measurement a {\it fixed} distance $R$ away from us,
the time interval $dt$ corresponding to any measurable (non-zero) value of $ds$
must go to infinity as $R\rightarrow R_{\rm h}$. (In the context of black-hole
physics, we recognize this effect as the divergent gravitational redshift
measured by a static observer outside the event horizon.) This result is generic
to all cosmologies though, as we shall see, in some cases $R_{\rm h}$ lies
beyond the distance $ct_0$ light has traveled since the big bang and is therefore
not an observable quantity. What does change from case to case, however, are the
constituents of the Universe, which directly determine $w$, and therefore the
function $f$ and the horizon's radius $R_{\rm h}$. Finding these quantities
will be the goal of the next section.

But we can already see directly from the FRW equations how $R_{\rm h}$
should behave for any given value of $w$. It is straightforward to demonstrate\cite{M08}
from Equations~(2)--(4) that
\begin{equation}
\dot{R}_{\rm h}={3\over 2}(1+w)c\;.
\end{equation}
Thus, $R_{\rm h}$ is an increasing function of cosmic time $t$ for any cosmology
with $w>-1$. It is fixed only for de Sitter, in which $\rho$ is a cosmological
constant and $w=-1$. In addition, there is clearly a demarcation at
$w=-1/3$. When $w<-1/3$, $R_{\rm h}$ increases more slowly than lightspeed,
and therefore our universe would be delimited by this horizon because light
would have traveled a distance $ct_0$ greater than $R_{\rm h}(t_0)$ since
the big bang. On the other hand, $R_{\rm h}$ is always greater than $ct$
when $w>-1/3$, and our observational limit would then simply be set
by the light travel distance $ct_0$.

\section{Specific Cosmologies}

\subsection{The De Sitter Universe $(w=-1)$}
As we pointed out above, de Sitter
himself published the earliest form of his metric using the coordinate $R$, 
though he was apparently not aware of its full significance. But as a simple,
initial application of the method we derived in \S~2 above, let us repeat this
calculation by considering a universe containing only a cosmological constant. In
this case,
\begin{equation}
a(t)=e^{H_0 t}\;,
\end{equation}
for which
\begin{equation}
ds^2=c^2\, dt^2- e^{2H_0t} \, \left[ dr^2+r^2\, d\Omega^2\right] \;.
\end{equation}
Clearly,
\begin{equation}
f(t)=\ln[a(t)]=H_0 t\;,
\end{equation}
so
\begin{equation}
\dot{f}=H_0\;.
\end{equation}
Thus, inserting this form of $\dot f$ into Equation~(16), we obtain
\begin{equation}
R_{\rm h}=c/H_0\;. 
\end{equation}

As we saw in Equation~(24), the de Sitter metric is unique among the various 
cosmologies in that the density $\rho$ is constant, and therefore the radius 
$R_{\rm  h}$ is fixed for all cosmic time $t$. However, for all
the other cosmologies we will consider below, $w>-1$, so $\dot{R}_{\rm h }>0$.
This means, of course, that whereas an observer can choose radii $R$ that are
always smaller than $R_{\rm h}$ in de Sitter, the same is not true when
$\dot{R}_{\rm h}>0$, since for any given $R$, $R_{\rm h}<R$ when $t\rightarrow 0$.

An important consequence of this distinction is that the interval $ds$ is
then imaginary at early times (since $R_{\rm h}<R$) for all but the de Sitter 
metric, and $dT$ can therefore not be integrated starting at $t=0$. But for de 
Sitter, we see from Equation~(18) (since $R_{\rm h}$ is independent of time) that
\begin{equation}
\Phi^{1/2}\,c\,dT\equiv\Phi^{1/2}\left[c\,dt+\left({R\over R_{\rm h}}\right)\Phi^{-1}\,dR\right]\;.
\end{equation}
A comparison with Equation~(19) tells immediately that the integrating factor $\eta(t,R)$ 
in the case of de Sitter is simply equal to $\Phi^{-1}$, and a straightforward integration 
of Equation~(30) thus gives
\begin{equation}
T(t,R)=t-{1\over 2H_0}\,\ln\Phi\;.
\end{equation}
At the origin, $T$ is of course equal to $t$, but since the gravitationally-induced
dilation increases with $R$, $T$ also includes the additional redshift seen at
progressively greater distances from the observer. Note, however, that for 
$dR=d\Omega=0$, $dT=dt$, so both of these time intervals diverge for a
finite $ds$ as $R\rightarrow R_{\rm h}$.

Written in terms of $T$ and $R$, the de Sitter metric becomes
\begin{equation}
ds^2=\Phi\,c^2\,dT^2-\Phi^{-1}\,dR^2-R^2\,d\Omega^2\;,
\end{equation}
which is the form originally presented by de Sitter.\cite{d17}

\subsection{A Cosmology with $R_{\rm h}(t)=ct$ {\rm (i.e.,} $w=-1/3$)}
This is the case in which $\dot{R}_{\rm h}=c$ in Equation
(24). Aside from being one of the simplest models of the universe that one
can construct from the FRW equations, it is also motivated by the observed
fact that $R_{\rm h}(t_0)\approx ct_0$ at the present time. If $w$ is
always $-1/3$, then it would not be surprising to see $R_{\rm h}(t_0)
=ct_0$ now, since this condition would have been true from the beginning.

But though this may seem like a highly idealized model that bears minimal
relevance to the real universe, it is actually singularly significant\cite{M07,M08}
because an equation of state
$w=-1/3$ is the only one for which the current age, $t_0$, of the
universe can equal the light-crossing time, $t_{\rm h}\equiv R_{\rm h}/c$.
For any other cosmology with $w<-1/3$, $t_0$ must be greater than $R_{\rm h}/c$.
We shall return to this shortly. For now, let us consider the transformation
of coordinates when $w=-1/3$, for which
\begin{equation}
a(t)= H_0 t\;.
\end{equation}
Then,
\begin{equation}
f(t)=\ln[a(t)]=\ln(H_0t)\;,
\end{equation}
and
\begin{equation}
\dot{f}=\frac{1}{t}\;.
\end{equation}

We therefore confirm that $R_{\rm h}=ct$ in this case, and from Equation~(18),
we obtain the metric
\begin{equation}
ds^2=\Phi\left[c\,dt+\left({R\over ct}\right)\Phi^{-1}\,dR\right]^{2} -
\Phi^{-1}{dR^2}-R^2\,d\Omega^2\;,
\end{equation}
with
\begin{equation}
\Phi=1-\left({R\over ct}\right)^2\;.
\end{equation}
As was the case with de Sitter, the cosmic time $dt$ diverges for a
measurable line element as $R\rightarrow R_{\rm  h}$ (which, however, is
equal to $ct$ here).

We may also write the metric in terms of $R$ and $T$. A dimensionless solution
to Equation~(22) is 
\begin{equation}
\eta(t,R) = \exp\left\{{1\over 2}\left({R\over ct}\right)^2\right\}\;,
\end{equation}
which leads to the metric
\begin{equation}
ds^2={e^{-(R/R_{\rm h})^2}}\Phi^{-1}\, c^2\,dT^2 -\Phi^{-1}\,{dR^2}-R^2\,d\Omega^2\;.
\end{equation}
But as we pointed out above, this form of the metric has only a limited applicability,
given that $ds^2<0$ when $R_{\rm h}<R$ and $dR=d\Omega=0$.

\subsection{Radiation Dominated Universe ($w=+1/3$)}
The real Universe may indeed contain a cosmological constant, in which case
it approaches a de Sitter spacetime asymptotically, given that its other
constituents have a density $\rho$ that drops off as $a(t)$ increases.
(We will revisit this below.) But let us now move beyond the simple de Sitter
application, and the special case with $w=-1/3$, and consider other cosmological
phases (as we currently understand them) that may have emerged since the big bang.

In the very beginning, when radiation dominated the equation of state (with
$w=w_{\rm rad}\equiv+1/3$), the expansion parameter was given as
\begin{equation}
a(t)=(2H_0 t)^{1/2}\;.
\end{equation}
In that case,
\begin{equation}
f(t)=\ln[a(t)]={1\over 2}\ln(2H_0 t)\;,
\end{equation}
and
\begin{equation}
\dot{f}=\frac{1}{2t}\;.
\end{equation}
Therefore $R_{\rm h}=2ct$. In this case, the visible Universe (extending out to $ct$)
never reaches $R_{\rm h}$, and the metric is
\begin{equation}
ds^2=\Phi\left[c\,dt+\left({R\over 2ct}\right)\Phi^{-1}\,dR\right]^{2} -
\Phi^{-1}{dR^2}-R^2\,d\Omega^2\;,
\end{equation}
with
\begin{equation}
\Phi=1-\left({R\over 2ct}\right)^2\;.
\end{equation}
Thus, measurements made at a fixed $R$ and $\Omega$ still produce a gravitationally-induced
dilation of $dt$ as $R$ increases, but this effect never becomes divergent within that 
portion of the Universe (i.e., within $ct_0$) that remains observable since the big bang.

For completeness, we note that a dimensionless solution for $\eta(t,R)$ from Equation~(22) is
\begin{equation}
\eta(t,R) = 1\;,
\end{equation}
and the metric written in terms of $R$ and $T$ is therefore 
\begin{equation}
ds^2= \Phi^{-1}\,c^2\, dT^2 - \Phi^{-1}\,dR^2 -R^2\,d\Omega^2\;.
\end{equation}
The same restrictions as before pertain to $T$ since here also $R_{\rm h}<R$
at early times, which would make $ds$ imaginary for $dR=d\Omega=0$ back then.

\subsection{Matter Dominated Universe ($w=0$)}
When $\rho$ is dominated by matter with $w\approx
w_{\rm matter}\equiv 0$, the expansion factor grows according to
\begin{equation}
a(t)=\left({3\over 2}H_0 t\right)^{2/3}\;.
\end{equation}
Therefore,
\begin{equation}
f(t)=\ln[a(t)]={2\over 3} \, \ln\left({3\over 2}H_0 t\right)\;,
\end{equation}
and
\begin{equation}
\dot{f}=\frac{2}{3t}\;.
\end{equation}
In this case, the radius of the cosmic horizon is $R_{\rm h}=(3/2)ct$,
and the metric becomes
\begin{equation}
ds^2=\Phi\left[c\,dt+\left({R\over 3ct/2}\right)\Phi^{-1}\,dR\right]^{2} -
\Phi^{-1}{dR^2}-R^2\,d\Omega^2\;,
\end{equation}
with
\begin{equation}
\Phi=1-\left({R\over 3ct/2}\right)^2\;.
\end{equation}
The situation is similar to that for a radiation dominated universe, in that
$R_{\rm h}$ always recedes from us faster than lightspeed (see Equation~24).
Although dilation is evident with increasing $R$, curvature alone does not
produce a divergent redshift.

A dimensionless integrating factor $\eta(t,R)$ is
\begin{equation}
\eta(t,R) = \sqrt{1+\frac{1}{2}\,\left(\frac{R}{R_{\rm h}} \right)^2 }\;.
\end{equation}
Thus, in terms of $R$ and $T$, 
\begin{equation}
ds^2= \Phi^{-1}\left[1+\frac{1}{2}\left(\frac{R}{R_{\rm h}}\right)^2 \right]^{-1}\,c^2\, dT^2
-\Phi^{-1}\,dR^2-R^2\,d\Omega^2\;.
\label{matterfinal}
\end{equation}

\subsection{Universe Delimited by the Cosmic Horizon ($-1<w<-1/3$)}
In general, $R_{\rm h}=(3/2)(1+w)ct$ for a universe with constant $w$. Except
for the special cases $w=-1$ and $w=-1/3$, the metric may therefore be written
\begin{equation}
ds^2=\Phi\left[c\,dt+\left({2R\over 3(1+w)ct}\right)\Phi^{-1}\,dR\right]^{2} -
\Phi^{-1}{dR^2}-R^2\,d\Omega^2\;,
\end{equation}
with
\begin{equation}
\Phi=1-\left({2R\over 3(1+w)ct}\right)^2\;.
\end{equation}
(Actually, both this form of the metric, as well as that given below in Equation~57, are
valid for constant values of $w$ greater than $-1/3$. However, in this section we
are primarily interested in equations of state for which $\dot{R}<c$, i.e., $w<-1/3$.) 
A cosmology with $w<-1/3$ is therefore delimited by the horizon in the sense that our
measurements---extending over a distance $ct_0$---would have probed regions of
spacetime that include $R_{\rm h}(t_0)$ today. 

For example, in what may be a close approximation to the Universe we have
currently (see next subsection), the line element for $w=-2/3$ is
\begin{equation}
ds^2=\Phi\left[c\,dt+\left({2R\over ct}\right)\Phi^{-1}\,dR\right]^{2} -
\Phi^{-1}{dR^2}-R^2\,d\Omega^2\;,
\end{equation}
where $\Phi=1-({2R/ ct})^2$. In this Universe, the cosmic horizon would lie
at a radius from us only half of the total distance traversed by light
since the big bang, and we would therefore be sampling phenomena near
the gravitational limit.

It is not difficult to see that the general form of the metric written in 
terms of $R$ and $T$ is
\begin{equation}
ds^2=\Phi^{-1}\,\left[1+{1\over 2}(1+3w)\left({R\over R_{\rm h}}\right)^2\right]^{(3w-1)/(3w+1)}
\,c^2\,dT^2-\Phi^{-1}\,dR^2-R^2\,d\Omega^2\;.
\end{equation}

\subsection{The $\Lambda$CDM Model}
Let us now examine the role played by $R_{\rm h}$ in the ``standard" model
of cosmology. Following the radiation-dominated era, the universal expansion
is believed to have been driven by a combination of matter plus a cosmological
constant, the latter emerging at later times since $\rho_{\rm matter}\propto a(t)^{-3}$,
whereas $\rho_\Lambda=$ constant. Based on the solutions we have examined thus far,
we anticipate that this kind of Universe does become $R_{\rm h}$-delimited, but
only after the cosmological constant dominates.

From the FRW equations with $\rho=\rho_{\rm matter}
+\rho_\Lambda$ and $w=w_{\rm matter}+w_\Lambda=-1$, where
$w_{\rm matter}=0$ and $w_\Lambda=-1$, we infer that
\begin{equation}
 a(t)=A\,\sinh^{2/3}\left(\frac{t}{t_{\Lambda}}\right)\;.
\end{equation}
The ratio $\rho_{\rm matter}/\rho_\Lambda$ is not known a priori;
its possible range of values is subsumed into the (time) constant
$t_\Lambda$. The greater the value of $\rho_{\rm matter}/\rho_\Lambda$,
the greater the value of $t_\Lambda$, pushing the transition from a
matter-dominated universe to a $\Lambda$-dominated one farther into
the future. Notice from Equation (58) that $a(t)\rightarrow {\rm constant}\times
t^{2/3}$ for $t\ll t_\Lambda$, in agreement with Equation~(47) for a
matter-dominated universe. At the other extreme, $a(t)\rightarrow {\rm constant}\times
\exp(2t/3t_\Lambda)$, the correct behavior exhibited in Equation~(25) for a
de Sitter universe. Clearly then,
\begin{equation}
t_\Lambda \equiv {2\over 3H_\infty}\;,
\end{equation}
where $H_\infty\equiv\lim_{t\rightarrow\infty} H(t)$ is the asymptotic (constant)
value of the Hubble constant describing a universe settled into its cosmological-constant
driven expansion.

Thus, in $\Lambda$CDM cosmology,
\begin{equation}
f(t)=\ln[a(t)]=\ln(A) +\frac{2}{3} \, \ln\left[\sinh (t/t_{\Lambda})\right]\;,
\end{equation}
and
\begin{equation}
\dot{f}= \frac{2}{3t_{\Lambda}\tanh(t/t_{\Lambda})}
= {H_\infty\over \tanh(3tH_\infty/2)}\;.
\end{equation}
It is trivial to check that $\dot{f}$ has the correct limiting forms
given in Equations~(28) (when $t\rightarrow\infty$) and (49)
(when $t\rightarrow 0$). In this case,
\begin{equation}
R_{\rm h}=\left({c\over H_\infty}\right)\tanh\left({3\over 2}tH_\infty\right)\;,
\end{equation}
and therefore
\begin{equation}
\dot{R}_{\rm h}={3\over 2}\left[1-\tanh^2\left({3\over 2}tH_\infty\right)\right]c\;.
\end{equation}
A direct comparison of this equation for $\dot{R}_{\rm h}$ with that given in Equation (24)
is not quite legitimate since $w$ was assumed to be constant in the earlier expression.
However, for simple analysis over short intervals of time $\Delta t\ll t$, it can still
be useful to approximate $w$ as follows:
\begin{equation}
w\approx-\tanh^2\left({3\over 2}tH_\infty\right)\;,
\end{equation}
which again has the correct behavior in the appropriate time limits, i.e.,
$w\rightarrow 0$ when $t\rightarrow 0$ and $w\rightarrow -1$ when $t\rightarrow\infty$.

In this cosmology, the observer does not experience
a divergent redshift with increasing $R$ at early times, since $\dot{R}_{\rm h}>c$ then,
but he will begin to encounter an observational limit at a finite radius $\sim R_{\rm h}$
after a ``transition" time $t_{\rm trans}$ estimated from the condition
\begin{equation}
ct_{\rm trans}=R_{\rm h}(t_{\rm trans})\;.
\end{equation}
Equation~(65) has two solutions, the trivial value $t_{\rm trans}=0$ (which is
probably irrelevant since the Universe is radiation dominated at the beginning),
and $t_{\rm trans}\sim 0.86/H_\infty$ (which is also approximately $t_\Lambda$).
In a sense, $t_{\rm trans}$ is roughly the point at which the Universe transitions
from being matter-dominated to $\Lambda$-dominated. Eventually, the Universe
becomes de Sitter and therefore $\dot{R}_{\rm h}\rightarrow 0$, with $R_{\rm h}$
settling at the fixed value $c/H_\infty$.

\section{Conclusions}
An observer measuring intervals of time at progressively greater distances from
his origin of coordinates sees a gravitationally-induced time dilation due to the
mass-energy content of the Universe between himself and the radius at which he
is making the observation. This gravitational redshift is manifested via the
appearance of the radius $R_{\rm h}$ (of our cosmic horizon) in the $g_{00}$
coefficient of the metric for all the cases we considered in this paper. Its
modification to the intervals we measure at radii $R>0$ is a generic feature
of any universe with $\rho\not=0$.

However, the actual form of $R_{\rm h}$, and its time derivative $\dot{R}_{\rm h}$,
depend on the equation of state $w\equiv p/\rho$. The correspondence between
$R_{\rm h}$ and the event horizon introduced earlier\cite{R56} was
discussed at length previously.\cite{M07} Briefly, $R_{\rm h}$ is an instantaneous
horizon that (except in the case $w=-1$) increases with cosmic time $t$. It
asymptotically approaches Rindler's (fixed) event horizon in situations where
$\dot{R}_{\rm h}<c$, and is equal to the latter in the case of de Sitter, for
which $R_{\rm h}$ is a constant.  In cosmologies with $w>-1/3$, $R_{\rm h}$ is
always greater than the distance traveled by light since the big bang, and therefore
does not represent a limit to our range of observations. Of course, in these cases
Rindler's event horizon also does not exist.

During the radiation- and matter-dominated eras, $\dot{R}_{\rm h}>c$, and though
measured time intervals $dt$ are gravitationally dilated relative to the values they
would otherwise have, they never diverge. In these cases, there is therefore
no physical limitation to how far we can see except, of course, as restricted by
the distance $ct$ light has traveled since $t=0$.

But for any universe with an equation of state $w<-1/3$, we find
that $\dot{R}_{\rm h}<c$, so there exists a {\it finite} radius
$\sim R_{\rm h}(t_0)< ct_0$ (where $t_0$ is the current age of the
Universe) at which $dt\rightarrow \infty$ for any finite interval $ds$.
Thus, although $R_{\rm h}$ increases with time $t$ (except in the
special case $w=-1$), our past lightcone is always truncated by
gravitational curvature.

In earlier work,\cite{M07,M08} we explored several observational consequences of
this phenomenon. Of particular interest is the peculiar observation that $R_{\rm h}(t_0)$
is currently equal (or nearly equal) to $ct_0$. As we have seen through the various
cosmological spacetimes we considered in this paper, one would expect the condition
$R_{\rm h}=ct$ to be met only if $w=-1/3$, in which case it would always be true.
Alternatively, it would be met just once, at a time $t_{\rm trans}\sim 0.86/H_\infty$
in the context of $\Lambda$CDM. For all other times $t>t_{\rm trans}$, $R_{\rm h}$ must
be less than $ct$. A more extended discussion of the physical interpretation of these
results has appeared elsewhere.\cite{M08}

Insofar as the nature of dark energy is concerned, however, the observed equality
(or near equality) of $R_{\rm h}(t_0)$ and $ct_0$ presents a problem for models in which
$w\approx -1$. Through the various spacetimes we have considered in this paper,
we see that the condition $R_{\rm h}=ct$ can only be attained for $w<-1/3$; otherwise,
$R_{\rm h}$ always exceeds $ct$. But the fact that $R_{\rm h}$ would then equal $ct$ just
once in the entire history of the universe if $w\not=-1/3$ constitutes an unacceptably
improbable coincidence that we should be seeing this transition occurring right now. We
therefore affirm our earlier conclusion that dark energy is probably not a cosmological
constant.

\section*{Acknowledgments}

This research was partially supported by NSF grant 0402502 at the
University of Arizona. Many inspirational discussions with Roy Kerr
are greatly appreciated. Part of this work was carried out at
Melbourne University and at the Center for Particle Astrophysics
and Cosmology in Paris.

\end{document}